\documentclass[pra, showpacs, twocolumn, floatfix]{revtex4}
\usepackage{graphicx}
\usepackage{epstopdf}
\usepackage{amsmath, amsfonts, amssymb, bm}
\begin{document}
\title{Phase dependence of the unnormalized second-order photon correlation function}
\author{Viorel \surname{Ciornea}}
\author{Profirie \surname{Bardetski}}
\author{Mihai A. \surname{Macovei}}
\email{macovei@phys.asm.md}
\affiliation{Institute of Applied Physics, Academy of Sciences of Moldova, 
Academiei str. 5, MD-2028 Chi\c{s}in\u{a}u, Moldova}
\date{\today}
\begin{abstract}
We investigate the resonant quantum dynamics of a multi-qubit ensemble in a microcavity. 
Both the quantum-dot subsystem and the microcavity mode are pumped coherently.
We found that the microcavity photon statistics depends on the phase difference 
of the driving lasers which is not the case for the photon intensity at resonant driving.
This way, one can manipulate the two-photon correlations. In particular, higher degrees 
of photon correlations and, eventually, stronger intensities are obtained. Furthermore, 
the microcavity photon statistics exhibits steady-state oscillatory behaviors as well as asymmetries.
\end{abstract}
\maketitle
\section{Introduction}
Quantum dots or {\it artificial atoms} can have sharp optical transitions,
similar to those of real atoms \cite{qd}. Applying a coherent laser field, one 
can address those transitions, emphasizing in particular, a two-level 
system. This way, a number of effects can be obtained and some of them
are known from pumping of real atoms with coherent laser fields. 
Particularly, resonance fluorescence from a coherently driven semiconductor 
quantum dot in a cavity was experimentally investigated in \cite{rf_exp}.
The observation of the Mollow triplet \cite{ml} from a quantum dot system 
was reported in \cite{mt1,mt2}. Dephasing of triplet-sideband optical emission 
of a resonantly driven InAs/GaAs quantum dot inside a microcavity was studied 
in Ref.~\cite{deph}. Furthermore, cascaded single-photon emission from the 
Mollow triplet sidebands of a quantum dot as well as spectral photon 
correlations were obtained in \cite{cr_exp}. A pronounced interaction between 
the quantum dot and the cavity has been observed even for detunings of many 
cavity linewidths \cite{off_r}. In the small Rabi frequency regime, 
subnatural linewidth single photons from a quantum dot were obtained too, 
in \cite{sm_rabi}. Moreover, self-homodyne measurement of a dynamic Mollow
triplet in the solid state systems was performed recently \cite{hom}.

When two or more quantum dots are close to each other on the emission wavelength 
scale then collective interactions come into play \cite{Puri,chk,ficek,kmek,kilin,jorg}. 
In particular, superradiance in an ensemble of quantum dots was experimentally 
observed in \cite{supr_exp} while the dynamics of quantum dot superradiance was 
investigated in \cite{dsup}. The collective fluorescence and decoherence of a few nearly 
identical quantum dots and superbunched photons via a strongly pumped near-equispaced 
multi-particle system were analyzed in \cite{clf1} and \cite{clf2}, respectively. Dicke states 
in multiple quantum dots systems were discussed too, in Ref.~\cite{sitek}. Furthermore, 
sub- and super-radiance phenomena in quantum dot nanolasers were investigated in \cite{prp}.
The collective modes of quantum dot ensembles in microcavities 
were obtained as well \cite{colm}. Finally, entanglement of two quantum dots was 
investigated in Ref.~\cite{gxl}. 

Here, we investigate the dynamics of a two-level quantum dot ensemble 
inside a microcavity. However, the developed approach applies to a real 
atomic sample as well. The microcavity mode together with the qubit 
subsystem are pumped with two distinct coherent electromagnetic fields. 
When the laser that resonantly pumps the qubit ensemble is moderately 
intense, i.e. the corresponding Rabi frequency is larger than the 
qubit-cavity coupling strength as well as the spontaneous and cavity 
decay rates, we found enhanced photon-photon correlations. In particular, 
the photon statistics displays oscillatory steady-state behaviors due to 
an interplay between the cavity and sponatenous emission decay rates.
Furthermore, the microcavity photon statistics depends on the phase 
difference of the applied coherent sources that can be a convenient 
mechanism to influence the second-order photon-photon correlations. 
An asymmetrical steady-state behavior of the second-order photon 
correlation function versus cavity-field detuning is observed due to 
the relative phase dependence. 

The article is organized as follows. In Section II we describe the analytical approach 
and the system of interest, and obtain the corresponding equations of motion. Section 
III deals with discussions of the obtained results. The Summary is given in Section IV.

\section{Quantum dynamics of a pumped multi-qubit system in a microcavity}
The Hamiltonian describing a wavelength-sized collection of pumped two-level artificial 
(or real) atomic system possessing the frequency $\omega_{0}$ and embedded in a 
microcavity of frequency $\omega_{c}$ is:
\begin{eqnarray}
H &=& \hbar\Delta a^{\dagger}a + \hbar g(a^{\dagger}S^{-}+aS^{+})+
\hbar \epsilon(a^{\dagger}e^{i\phi_{1}}+ae^{-i\phi_{1}}) \nonumber \\
&+& \hbar \Omega(S^{+}e^{i\phi_{2}}+S^{-}e^{-i\phi_{2}}).
\label{HM}
\end{eqnarray}
Here, both the atomic sample, and the microcavity mode, are interacting with coherent sources of 
frequency $\omega_{L1}=\omega_{L2}\equiv \omega_{L}$, in a frame rotating at $\omega_{L}$,
and we have assumed that $\omega_{0}=\omega_{L}$. In the Hamiltonian (\ref{HM})
the first term describes the cavity free energy with $\Delta=\omega_{c}-\omega_{L}$, 
while the second one characterize the interaction of the quantum dot system with 
the microcavity mode via the coupling $g$. The third term takes into account the 
interaction of the microcavity mode with the first coherent light source of 
amplitude $\epsilon$ and phase $\phi_{1}$. The last term considers the interaction 
of the qubit subsystem with the second laser with $\Omega$ and $\phi_{2}$ being the 
corresponding Rabi frequency and phase. The collective operators 
$S^{+}=\sum^{N}_{j=1}S^{+}_{j}$=$\sum^{N}_{j=1}|2\rangle_{j}{}_{j}\langle 1|$ and 
$S^{-}=[S^{+}]^{\dagger}$ obey the commutation relations for su(2) algebra: 
$[S^{+},S^{-}]=2S_{z}$ and $[S_{z},S^{\pm}]=\pm S^{\pm}$. Here 
$S_{z}=\sum^{N}_{j=1}S_{zj}=\sum^{N}_{j=1}(|2\rangle_{j}{}_{j}\langle 2|-
|1\rangle_{j}{}_{j}\langle 1|)/2$ is the bare-state inversion 
operator while $N$ is the number of quantum dots involved. $|2\rangle_{j}$ and 
$|1\rangle_{j}$ are the excited and ground state of the $j$th qubit, respectively.
Further, $a^{\dagger}$ and $a$ are the creation and the annihilation operator of the 
electromagnetic field (EMF), and satisfy the standard bosonic commutation relations, 
i.e., $[a,a^{\dagger}]=1$, and $[a,a]=[a^{\dagger},a^{\dagger}]=0$. We have supposed 
here that the quantum dot system couples to the laser and microcavity fields with the 
same coupling strength, i.e. the linear extension of the quantum dot ensemble is smaller 
than the relevant emission wavelength.

In what follows, we are interested in the laser dominated regime where $\Omega \gg \{g,\gamma,\kappa\}$
(here $\gamma$ and $\kappa$ are the spontaneous and cavity decay rates, respectively) and shall describe 
our system using the dressed-states formalism \cite{Puri,chkPRL}:
\begin{eqnarray}
|1\rangle_{j}=\frac{1}{\sqrt{2}}( |\bar 1\rangle_{j} + |\bar 2\rangle_{j}), ~~~
|2\rangle_{j}=\frac{1}{\sqrt{2}}( |\bar 2\rangle_{j} - |\bar 1\rangle_{j}).
\label{DS}
\end{eqnarray}
Before applying the transformation (\ref{DS}) we performed the substitution 
$a^{\dagger}e^{i\phi_{1}}=\tilde a^{\dagger}$ and $S^{+}e^{i\phi_{2}}=\tilde S^{+}$
and dropped the tilde afterwards. Restricting ourselves to values of $\Delta \ll \Omega$  
and secular approximation, one then arrives at the following master equation 
describing our system:
\begin{eqnarray}
\frac{d}{dt}\rho(t) &+& i[H_{0},\rho] = -\Gamma_{0}[R_{z},R_{z}\rho] 
-\Gamma\{[R^{+},R^{-}\rho] \nonumber \\
&+& [R^{-},R^{+}\rho]\} -\kappa[a^{\dagger},a\rho] + H.c..
\label{ME}
\end{eqnarray}
Here 
\begin{eqnarray*}
H_{0}= \Delta a^{\dagger}a + R_{z}(g^{\ast}_{0}a^{\dagger} + g_{0}a)+
\epsilon(a^{\dagger} + a),
\end{eqnarray*}
where $g_{0}=ge^{i\phi}/2$ and $g^{\ast}_{0}=ge^{-i\phi}/2$ with 
$\phi = \phi_{1}-\phi_{2}$. $\Gamma_{0} = \gamma/4$ and $\Gamma = (\gamma+\gamma_{d})/4$ 
with $2\gamma$ being the single-qubit spontaneous decay rate, while $\gamma_{d}$ is the 
quantum dot dephasing rate. The new quasispin operators,
i.e. $R^{+}=\sum^{N}_{j=1}|\bar 2\rangle_{j}{}_{j}\langle \bar 1|$, $R^{-}=[R^{+}]^{\dagger}$
and $R_{z}=\sum^{N}_{j=1}(|\bar 2\rangle_{j}{}_{j}\langle \bar 2|-|\bar 1\rangle_{j}{}_{j}\langle \bar 1|)$
are operating in the dressed state picture. They obey the following commutation relations: 
$[R^{+},R^{-}]=R_{z}$ and $[R_{z},R^{\pm}]=\pm 2R^{\pm}$. Notice the dependence of the coupling 
strength $g_{0}$ on the phase difference of the applied coherent sources. Additional and different phase 
dependent effects can be found in \cite{ficek,kmek}.

In the next subsection, we shall obtain the equations of motion of variables of interest 
in order to calculate the second-order microcavity photon correlation function: 
$g^{(2)}(0)=\langle a^{\dagger}a^{\dagger}aa\rangle/(\langle a^{\dagger}a\rangle)^{2}$.
Values of $g^{(2)}(0)$ smaller than unity describe sub-Poissonian photon statistics and
it is a quantum effect. Poissonian photon-statistics has $g^{(2)}(0)=1$. $g^{(2)}(0)>1$ 
characterizes super-Poissonian photon statistics. In particular for thermal light one has
$g^{(2)}(0)=2$ and, therefore, we are interested in correlations larger than two, i.e. 
$g^{(2)}(0)>2$.

\subsection{Equations of motion}
Using Eq.~(\ref{ME}) one can obtain the following equations of motion 
in order to calculate the microcavity photon intensity and their 
second-order photon-photon correlations:
\begin{eqnarray}
&\frac{d}{dt}\langle a^{\dagger}a\rangle =i\epsilon(\langle a\rangle-\langle a^{\dagger}\rangle) 
+ ig_{0}\langle R_{z}a\rangle - ig^{\ast}_{0}\langle R_{z}a^{\dagger}\rangle \nonumber \\
&- 2\kappa\langle a^{\dagger}a\rangle, \nonumber \\
&\frac{d}{dt}\langle a^{\dagger}\rangle =i\epsilon + ig_{0}\langle R_{z}\rangle - (\kappa-i\Delta)\langle a^{\dagger}\rangle, \nonumber \\
&\frac{d}{dt}\langle R_{z}a\rangle =-i\epsilon \langle R_{z}\rangle - ig^{\ast}_{0}\langle R^{2}_{z}\rangle 
- (4\Gamma + \kappa +i\Delta)\langle R_{z}a\rangle, \nonumber \\
&\frac{d}{dt}\langle a^{\dagger^2}a^{2}\rangle =2i\epsilon(\langle a^{\dagger}a^{2}\rangle-\langle a^{\dagger^2}a\rangle) 
+ 2ig_{0}\langle R_{z}a^{\dagger}a^{2}\rangle \nonumber \\
&- 2ig^{\ast}_{0}\langle R_{z}a^{\dagger^2}a\rangle - 4\kappa\langle a^{\dagger^2}a^{2}\rangle, \nonumber \\
&\frac{d}{dt}\langle R_{z}a^{\dagger}a^{2}\rangle =i\epsilon(\langle R_{z}a^{2}\rangle-2\langle R_{z}a^{\dagger}a\rangle) 
+ ig_{0}\langle R^{2}_{z}a^{2}\rangle \nonumber \\
&- 2ig^{\ast}_{0}\langle R^{2}_{z}a^{\dagger}a\rangle - (3\kappa+4\Gamma+i\Delta) \langle R_{z}a^{\dagger}a^{2}\rangle, \nonumber \\
&\frac{d}{dt}\langle a^{\dagger}a^{2}\rangle =i\epsilon(\langle a^{2}\rangle-2\langle a^{\dagger}a\rangle) 
+ ig_{0}\langle R_{z}a^{2}\rangle - 2ig^{\ast}_{0}\langle R_{z}a^{\dagger}a\rangle \nonumber \\
&- (3\kappa + i\Delta) \langle a^{\dagger}a^{2}\rangle, \nonumber \\
&\frac{d}{dt}\langle R^{2}_{z}a^{2}\rangle =-2i\epsilon\langle aR^{2}_{z}\rangle 
- 2ig^{\ast}_{0}\langle R^{3}_{z}a\rangle + 16\Gamma j(j+1)\langle a^{2}\rangle\nonumber \\
&- (2\kappa+12\Gamma+2i\Delta) \langle a^{2}R^{2}_{z}\rangle, \nonumber \\
&\frac{d}{dt}\langle R^{2}_{z}a^{\dagger}a\rangle =i\epsilon (\langle aR^{2}_{z}\rangle -\langle a^{\dagger}R^{2}_{z}\rangle)
+ ig_{0}\langle R^{3}_{z}a\rangle - ig^{\ast}_{0}\langle R^{3}_{z}a^{\dagger}\rangle \nonumber \\
&+ 16\Gamma j(j+1)\langle a^{\dagger}a\rangle - (2\kappa+12\Gamma)\langle a^{\dagger}aR^{2}_{z}\rangle, \nonumber \\
&\frac{d}{dt}\langle R^{3}_{z}a\rangle =-i\epsilon \langle R^{3}_{z}\rangle - ig^{\ast}_{0}\langle R^{4}_{z}\rangle 
- (24\Gamma + \kappa +i\Delta)\langle R^{3}_{z}a\rangle \nonumber \\
&+16\Gamma(3j(j+1)-1)\langle aR_{z}\rangle, \nonumber \\
&\frac{d}{dt}\langle a^{2}\rangle =-2i\epsilon \langle a\rangle - 2ig^{\ast}_{0}\langle R_{z}a\rangle -
(2\kappa+2i\Delta)\langle a^{2}\rangle, \nonumber \\
&\frac{d}{dt}\langle R_{z}a^{2}\rangle =-2i\epsilon \langle R_{z}a\rangle 
- 2ig^{\ast}_{0}\langle R^{2}_{z}a\rangle \nonumber \\
&-(2\kappa+4\Gamma+2i\Delta)\langle R_{z}a^{2}\rangle, \nonumber \\
&\frac{d}{dt}\langle R_{z}a^{\dagger}a\rangle =i\epsilon (\langle aR_{z}\rangle -\langle a^{\dagger}R_{z}\rangle)
+ ig_{0}\langle R^{2}_{z}a\rangle - ig^{\ast}_{0}\langle R^{2}_{z}a^{\dagger}\rangle \nonumber \\
&- (2\kappa+4\Gamma)\langle a^{\dagger}aR_{z}\rangle, \nonumber \\
&\frac{d}{dt}\langle R^{2}_{z}a\rangle =-i\epsilon \langle R^{2}_{z}\rangle - ig^{\ast}_{0}\langle R^{3}_{z}\rangle 
- (12\Gamma + \kappa +i\Delta)\langle R^{2}_{z}a\rangle \nonumber \\
&+16\Gamma j(j+1)\langle a\rangle.
\label{EqM}
\end{eqnarray}
The system of equations (\ref{EqM}) is not complete. Additional equations
are necessary for the qubit subsystem operators $\langle R_{z}\rangle$,
$\langle R^{2}_{z}\rangle$ etc. However, we shall represent the steady-state 
expectation values of the field correlators $\langle a^{\dagger}a\rangle$ 
and $\langle a^{\dagger^{2}}a^{2}\rangle$ via the quantum dot operators alone.
The expectation values of the quantum dot operators will be evaluated in a 
different way as described in the next subsection. Note that in deriving the 
above system of equations we have used the relation: 
$R^{2}_{z}/4+(R^{+}R^{-}+R^{-}R^{+})/2=j(j+1)$, where $j=N/2$.

\subsection{Qubit subsystem correlations}
As it was mentioned in the previous subsection, the steady-state values of 
field correlators as well as the qubit-field correlators can be expressed 
via the expectation values of the dressed-state inversion 
$\langle R^{n}_{z}\rangle$, $\{n \in 1,2,3,4\}$. These qubit-subsystem 
operators can be obtained from the master equation (\ref{ME}) by observing 
that any diagonal form of operators 
$R^{+m}R^{n}_{z}R^{-m}$, $\{ m,n \in 0, 1, \cdots \}$, 
commute with $H_{0}$. Therefore, the steady-state values of these operators 
are determined only by the dissipation part of the master equation. It is 
not difficult to show that the steady-state solution of the qubit subsystem 
master equation is \cite{Puri}: 
\begin{eqnarray}
\rho_{q}=\frac{\hat I}{N+1}, 
\label{rq}
\end{eqnarray}
where $\hat I$ is the unity operator. Consider an atomic 
coherent state $|n\rangle$, denoting a symmetrized $N$-atom 
state in which $N - n$ particles are in the lower dressed 
state $|\tilde 1\rangle$ and $n$ atoms are excited to the 
upper dressed state $|\tilde 2\rangle$. One can calculate 
the expectation values of any atomic correlators of interest 
using the relations: $R^{+}|n\rangle=\sqrt{(N-n)(n+1)}|n+1\rangle$,  
$R^{-}|n\rangle=\sqrt{n(N-n+1)}|n-1\rangle$, and 
$R_{z}|n\rangle=(2n-N)|n\rangle$. In particular, the steady-state 
expectation values of collective dressed-state inversion operator 
can be easily evaluated, namely:
\begin{eqnarray}
\langle R^{2}_{z}\rangle &=& \frac{N}{3}(N+2), \nonumber \\
\langle R^{4}_{z}\rangle &=& \frac{N}{15}(N+2)(3N^{2}+6N-4),
\label{rzn}
\end{eqnarray}
while $\langle R_{z}\rangle=\langle R^{3}_{z}\rangle=0$. 

In the following Section we shall discuss the microcavity photon statistics.
\begin{figure}[t]
\includegraphics[width=7cm]{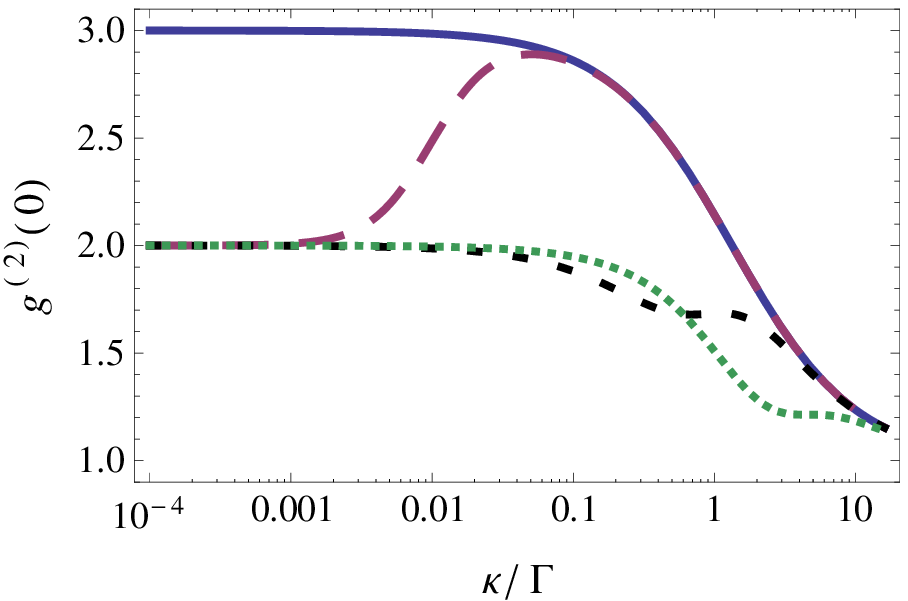}
\hspace{0.5cm}
\includegraphics[width=7cm]{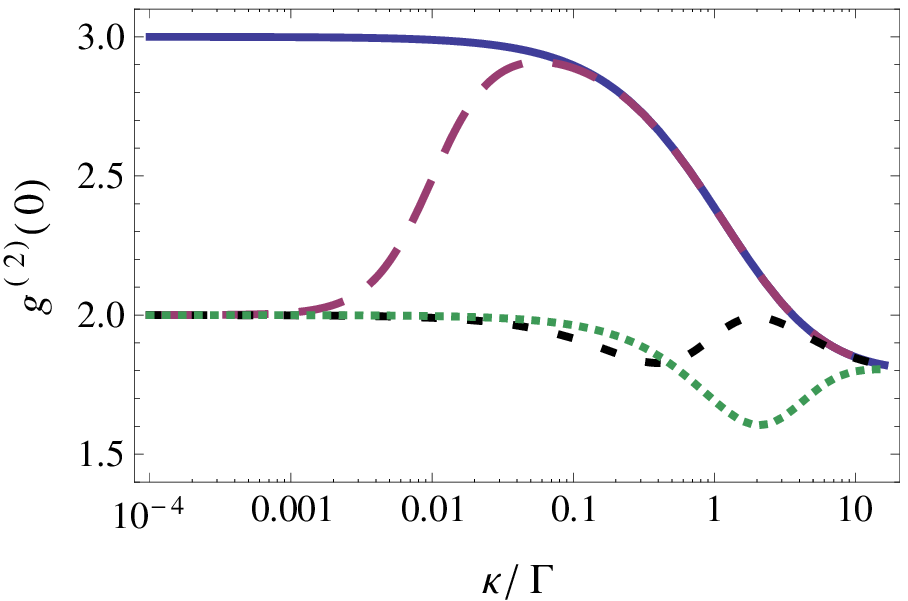}
\begin{picture}(0,0)
\put(-30,250){(a)}
\put(-30,110){(b)}
\end{picture}
\caption{\label{fig-1} The steady-state dependence of the microcavity second-order photon correlation 
function $g^{(2)}(0)$ as a function of $\kappa/\Gamma$ for $\epsilon=0$.
The solid line is for $\Delta/\Gamma=0$, the long-dashed line stands for 
$\Delta/\Gamma=0.01$, the short-dashed curve corresponds to $\Delta/\Gamma=1$ 
while the dotted line to $\Delta/\Gamma=4$. (a) N=1 while (b) N=20.}
\end{figure}
\begin{figure}[t]
\includegraphics[width=7cm]{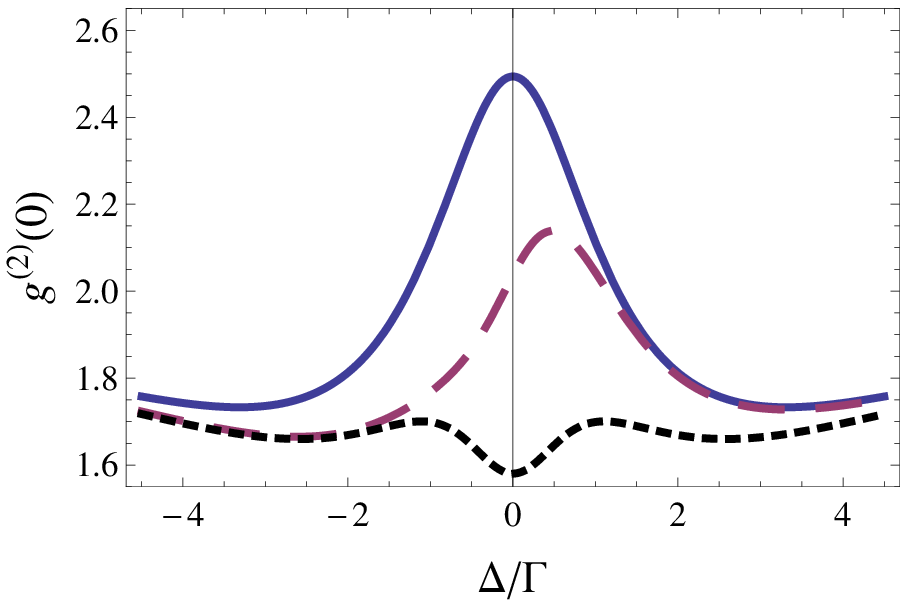}
\hspace{0.5cm}
\includegraphics[width=7cm]{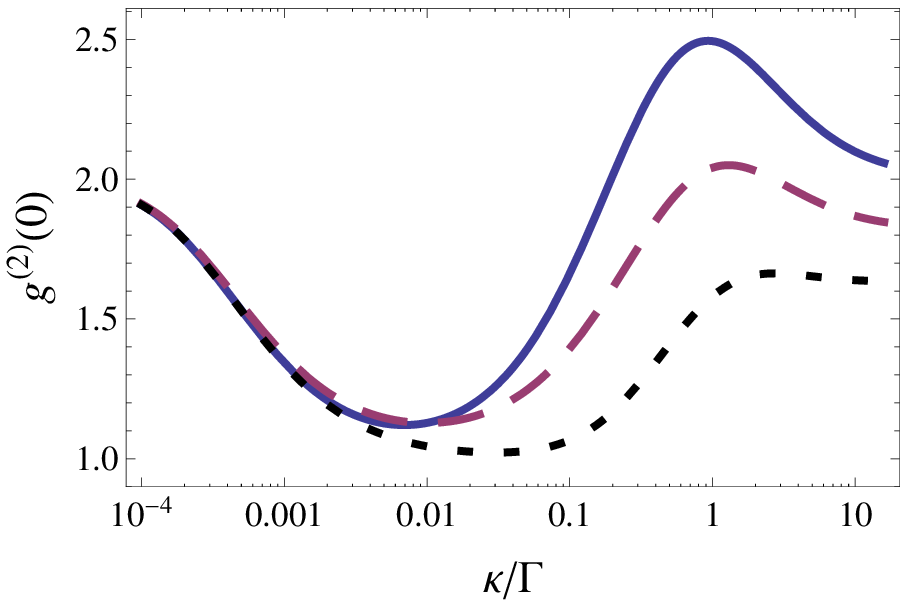}
\begin{picture}(0,0)
\put(-155,255){(a)}
\put(-155,105){(b)}
\end{picture}
\caption{\label{fig-2} The steady-state dependence of the microcavity  
second-order photon correlation function $g^{(2)}(0)$ as a function of (a) $\Delta/\Gamma$
and (b) $\kappa/\Gamma$.
The solid line is for $\phi=0$, the long-dashed line corresponds to
$\phi=\pi/4$ while the short-dashed curve to $\phi=\pi/2$. Here $\epsilon/\Gamma=20$, 
$g/\Gamma=10$, and $N=20$. Other parameters are: 
(a) $\kappa/\Gamma=1$ and (b) $\Delta/\Gamma=0.01$.}
\end{figure}

\section{Results and Discussion}
The general expression for the second-order photon correlation 
function is too cumbersome and, therefore, we shall represent it 
analytically for few particular cases. If $g_{0}=0$ one has $g^{(2)}(0)=1$ 
while when $\{\epsilon,\Delta\}=0$ we have:
\begin{eqnarray}
g^{(2)}(0) &=& \frac{3(4+\kappa/\Gamma)^{2}}{(4+3\kappa/\Gamma)(6+\kappa/\Gamma)\langle R^{2}_{z}\rangle^{2}}
\biggl\{ \frac{\langle R^{4}_{z}\rangle}{1+24\Gamma/\kappa} \nonumber \\
&+& \frac{8j(j+1)\langle R^{2}_{z}\rangle}{4+\kappa/\Gamma} 
+ \frac{16(3j(j+1)-1)\langle R^{2}_{z}\rangle}{(1+4\Gamma/\kappa)(24+\kappa/\Gamma)}\biggr\}. \nonumber \\
\label{g20}
\end{eqnarray}
One can observe here that the second-order correlation function does not 
depend on microcavity-dot coupling strength $g_{0}$. This is the case also for 
$\epsilon=0$ and $\Delta \not=0$. However, in general, i.e. when $\epsilon \not=0$, 
the microcavity photon correlation function depends on $g_{0}$. Note that the 
two-photon correlator $\langle a^{\dagger}a^{\dagger}aa\rangle \propto N^{4}$ while 
the photon intensity $\langle a^{\dagger}a\rangle \propto N^{2}$ and, thus, we have 
an enhancement of these correlations due to collectivity. 

In what follows, using the system of equations (\ref{EqM}) and for some particular 
cases the expression (\ref{g20}), we shall describe in details the microcavity 
second-order photon correlation function for various parameters of interest.
We proceed by considering that the microcavity mode is not additionally pumped, 
i.e. $\epsilon=0$. Figure (\ref{fig-1}) shows the dependence of the second-order 
correlation function as a function of $\kappa/\Gamma$ for various cavity detunings 
and number of quantum dots involved. At the exact resonance, that is $\Delta/\Gamma=0$, 
one can observe larger photon correlations, i.e. $g^{(2)}(0)=3$, while their intensity 
is being also enhanced due to collectivity. The picture is different for the 
off-resonance case. As long as $\kappa \ll \Delta \not=0$ the photon statistics is similar 
to that of a thermal light, i.e. $g^{(2)}(0)=2$. However, for intermediate detunings one can 
observe a oscillatory behavior of the second-order correlation function due to the interplay 
of $\kappa$ and $\Gamma$. As the detuning is further increased the two-photon correlation 
shows a dip because of the off-resonant driving (see Fig.~\ref{fig-1}). The Fig.~\ref{fig-1}(b) 
does not change if one further increase the number of quantum dots.

To further understand the steady-state behaviors of the photon-photon correlation function for $\epsilon \not=0$, 
in Figure~(\ref{fig-2}) we plot $g^{(2)}(0)$ again. However in this case, one observes a dependence of the normalized 
second-order correlation function $g^{(2)}(0)$ on the phase difference $\phi$ of the applied coherent sources. 
It is easy to show that the microcavity photon intensity (see Eqs.~\ref{EqM})
\begin{eqnarray}
\langle a^{\dagger}a\rangle = \frac{\epsilon^{2}}{\kappa^{2}+\Delta^{2}}
+ \frac{(\kappa+4\Gamma)|g_{0}|^{2}\langle R^{2}_{z}\rangle}
{[(\kappa+4\Gamma)^{2}+\Delta^{2}]\kappa}
\label{its}
\end{eqnarray}
does not depend on $\phi$ in this particular case. Therefore, the phase difference appears 
in the unnormalized second-order correlator $\langle a^{\dagger}a^{\dagger}aa\rangle$.
This happens due to feasibility of scattering two photons from different applied coherent 
sources giving rise to interferences, i.e., phase dependent effects. In Fig.~\ref{fig-2}(a) one 
can observe an asymmetrical steady-state behavior of the second-order correlation function for $\phi=\pi/4$. 
Furthermore, the maximum at $\Delta/\Gamma=0$ for $\phi=0$ turns into a minimum for $\phi=\pi/2$ 
(see the solid and short-dashed curves in Fig.~\ref{fig-2}a, respectively). Thus, the relative phase 
between the applied coherent sources can be a convenient tool to manipulate the photon statistics. 
Particularly, one can generate coherent light, i.e. $g^{(2)}(0)\approx 1$, despite of spontaneous incoherent photon 
scattering into the cavity mode (see Fig.~\ref{fig-2}b).  Again, the photon intensity as well as their second-order correlations 
are enhanced due to collectivity.

\section{Summary}
In summary, we have investigated the interaction of a collection of laser-pumped artificial atoms embedded in a leaking 
optical microcavity. Particularly, we were interested in photon statistics of the scattered photons into the cavity mode.
We have found that the photon statistics depends on the phase difference between the coherent sources pumping the 
quantum dot system and the cavity mode, respectively. Various steady-state behaviors of photon correlations were 
shown to occur.

\section*{Acknowledgment}
We acknowledge the financial support by the Academy of Sciences of Moldova, grant No. 15.817.02.09F. 



\begin{thebibliography}{33}
\bibitem{qd} C. Santori and Y. Yamamoto, Nature Phys. {\bf 5}, 173 (2009).

\bibitem{rf_exp} A. Muller, E. B. Flagg, P. Bianucci, X. Y. Wang, D. G. Deppe, W. Ma, 
J. Zhang, G. J. Salamo, M. Xiao and C. K. Shih, Phys. Rev. Lett. {\bf 99}, 187402 (2007).

\bibitem{ml} B. R. Mollow, Phys. Rev. {\bf 188}, 1969 (1969).

\bibitem{mt1} A. N. Vamivakas, Y. Zhao, C.-Y. Lu and M. Atat\"{u}re, Nature Phys. {\bf 5}, 198 (2009).

\bibitem{mt2} E. B. Flagg, A. Muller, J. W. Robertson, S. Founta, D. G. Deppe, M. Xiao, W. Ma, G. J. Salamo 
and C. K. Shih, Nature Phys. {\bf 5}, 203 (2009).

\bibitem{deph} S. M. Ulrich, S. Ates, S. Reitzenstein, A. L\"{o}ffler, A. Forchel and P. Michler, 
Phys. Rev. Lett. {\bf 106}, 247402 (2011).

\bibitem{cr_exp} A. Ulhaq, S. Weiler, S. M. Ulrich, R. Ro{\rm $\beta$}bach, M. Jetter and P. Michler,
Nature Photonics {\bf 6}, 238 (2012).

\bibitem{off_r} A. Majumdar, A. Papageorge, E. D. Kim, M. Bajcsy, H. Kim, P. Petroff and  J. Vuckovic,  
Phys. Rev. B {\bf 84}, 085310 (2011).

\bibitem{sm_rabi} C. Matthiesen, A. N. Vamivakas  and M. Atature,   Phys. Rev. Lett. {\bf 108}, 093602 (2012).

\bibitem{hom} K. A. Fischer, K. M\"{u}ller, A. Rundquist, T. Sarmiento, A. Y. Piggott,
Y. Kelaita, C. Dory, K. G. Lagoudakis and J. Vuckovic, Nature Photonics {\bf 10}, 163 (2016).

\bibitem{Puri} R. R. Puri {\it Mathematical Methods of Quantum Optics} (Springer, Berlin 2001), 
especially Chap. 12 and references therein.

\bibitem{chk} C. H. Keitel, M. O. Scully and G. S\"{u}ssmann, Phys. Rev. A {\bf 45}, 3242 (1992).

\bibitem{ficek} Z. Ficek and S. Swain {\it Quantum Interference and Coherence: Theory and Experiments} 
(Springer, Berlin, 2005).

\bibitem{kmek} M. Kiffner, M. Macovei, J. Evers and C. H. Keitel, Progress in Optics {\bf 55}, 85 (2010).

\bibitem{kilin} S. Ya. Kilin, Sov. Phys. JETP {\bf 51}, 1081 (1980).

\bibitem{jorg} P. Longo and J. Evers, Phys. Rev. Lett. {\bf 112}, 193601 (2014).

\bibitem{supr_exp} M. Scheibner, T. Schmidt, L. Worschech, A. Forchel, G. Bacher, T. Passow 
and D. Hommel, Nature Phys. {\bf 3}, 106 (2007).

\bibitem{dsup} V. I. Yukalov and E. P. Yukalova, Phys. Rev. B {\bf 81}, 075308 (2010).

\bibitem{clf1} A. Sitek and P. Machnikowski, Phys. Rev. B {\bf 75}, 035328 (2007).

\bibitem{clf2} M. Macovei and C. H. Keitel, Phys. Rev. B {\bf 75}, 245325 (2007).

\bibitem{sitek} A. Sitek and A. Manolescu, Phys. Rev. B {\bf 88}, 043807 (2013).

\bibitem{prp} H. A. M. Leymann, A. Foerster, F. Jahnke, J. Wiersig and C. Gies,  Phys. Rev. Applied {\bf 4}, 044018 (2015).

\bibitem{colm} N. S. Averkiev, M. M. Glazov and A. N. Poddubnyi, JETP {\bf 108},  836 (2009).

\bibitem{gxl} G.-x. L, Y.-p. Yang, K. Allaart and D. Lenstra, Phys. Rev. A {\bf 69},  014301 (2004).

\bibitem{chkPRL} C. H. Keitel, Phys. Rev. Lett. {\bf 83}, 1307 (1999).
\end{thebibliography}
\end{document}